\documentclass[pra,twocolumn,showpacs,amsmath,amssymb,floatfix,superscriptaddress]{revtex4}

\usepackage{graphicx} 
\usepackage{braket}   
\newcommand{\rme}{\mathrm{e}}
\newcommand{\rmi}{\mathrm{i}}
\newcommand{\rmd}{\mathrm{d}}
\newcommand{\rmk}{|\mathbf{k}|}
\newcommand{\rmR}{\mathrm{R}}
\newcommand{\rmA}{\mathrm{A}}
\newcommand{\bfk}{\mathbf{k}}
\newcommand{\bfr}{\mathbf{r}}
\newcommand{\bfR}{\mathbf{R}}
\newcommand{\bfv}{\mathbf{v}}

\newcommand{\bfA}{\mathbf{A}}
\newcommand{\bfF}{\mathbf{F}}
\newcommand{\bfn}{\mathbf{n}}

\begin{document}

\title{Two-center minima in harmonic spectra from aligned polar molecules}

\author{Adam Etches}
\affiliation{Lundbeck Foundation Theoretical Center for Quantum System Research, Department of Physics and Astronomy, Aarhus University, 8000 Aarhus C, Denmark}
\affiliation{Department of Physics and Astronomy, Louisiana State University, Baton Rouge, Louisiana 70803-4001, USA}

\author{Mette B. Gaarde}
\affiliation{Department of Physics and Astronomy, Louisiana State University, Baton Rouge, Louisiana 70803-4001, USA}

\author{Lars Bojer Madsen}
\email[]{bojer@phys.au.dk}
\affiliation{Lundbeck Foundation Theoretical Center for Quantum System Research, Department of Physics and Astronomy, Aarhus University, 8000 Aarhus C, Denmark}

\date{\today}

\begin{abstract}
  We extend a model of two-center interference to include the superposition of opposite orientations in aligned polar molecules. We show that the position of the minimum in the harmonic spectrum from both aligned and oriented CO depends strongly on the relative recombination strength at different atoms, not just the relative phase. We reinterpret the minimum in aligned CO as an interference between opposite orientations, and obtain good agreement with numerical calculations. Inclusion of the first-order Stark effect shifts the position of the interference minimum in aligned CO even though aligned molecules do not posses total permanent dipoles. We explain the shift in terms of an extra phase that the continuum electron of \emph{oriented} CO accumulates due to the Stark effect.
\end{abstract}


\maketitle

\section{Introduction}

Two-center interference minima are an important spectral signature in high-harmonic spectra from homonuclear molecules~\cite{lein:023805, lein:183903}. Two-center interference is often described within the framework of the three-step model, in which (i) an electron tunnels through the effective barrier formed by the molecular potential and a strong laser field, (ii) picks up kinetic energy in the continuum, and (iii) emits high-energy photons upon recombination into the ground state~\cite{krause:3535, corkum:1994}. The harmonic emission splits into contributions from recombination at each atomic center, with the possibility of destructive interference between different centers. The position of the minimum depends on the momentum of the returning continuum electron, and the projected internuclear distance~\cite{lein:023805, lein:183903}.

It has recently been shown within the Lewenstein model~\cite{lewenstein:2117} that two-center interference in oriented polar molecules is more complicated than in aligned homonuclear molecules. The added complication is caused by different intrinsic recombination phases at different atoms, which can shift the interference minimum~\cite{zhu:137866}. We show that different recombination strengths also strongly affect the minimum position, and extend the analysis to the case of aligned polar molecules, which have a well-defined alignment of the internuclear axis, but not a well-defined head-to-tail direction. Field-free molecular alignment is much easier to obtain experimentally than full orientation~\cite{stapelfeldt:543}. We obtain a simple equation for the prediction of the minimum position by reinterpreting the minimum as an interference between opposite orientations. Interference effects have also recently been considered in homonuclear~\cite{odzak:023414} and in heteronuclear~\cite{augstein:055601} diatomic molecules using a different model.

Polar molecules differ from homonuclear molecules in that their molecular orbitals have permanent dipole moments that interact strongly with electric fields. This requires that a model of high-order harmonic generation (HHG) from polar molecules should include the effect of the laser field on the ground state, something which is typically ignored for atomic and homonuclear systems. The response of the ground state to the laser field can be included to lowest order by adiabatically Stark shifting the energy levels of the field-free molecular orbitals to the instantaneous value of the electric field~\cite{etches:155602}. The influence of Stark shifts on strong-field ionization has recently been investigated in tunneling theory~\cite{holmegaard:428, dimitrovski:023405, hansen:023406, lein:ModOpt}, in the strong-field approximation~\cite{dimitrovski:053404}, and in calculations using the time-dependent Schr\"odinger equation in the single-active electron approximation~\cite{abu-samha:043413}. 

Our present calculations show that the Stark effect shifts the interference minimum in aligned CO by as much as three harmonic orders when using experimentally feasible pulse parameters \emph{even though aligned molecules do not posses total permanent dipoles}. The shift is explained in terms of the phase that the continuum electron of the \emph{oriented} molecule accumulates from the Stark effect. Our results indicate that inclusion of the Stark effect is important when modelling spectral features of HHG from polar molecules.

The paper is organized as follows. We briefly describe how we calculate HHG spectra in Sec.~\ref{calculating spectra}, then present our model of two-center interference from aligned molecules in Sec.~\ref{aligned molecules}. In Sec.~\ref{Stark shift} we discuss how the Stark effect shifts the position of the minimum, and explain the shift in terms of the Stark phase accumulated by an electron in the continuum. We summarize our results in Sec.~\ref{conclusions}. Two derivations are relegated to Appendix~\ref{derivations}.

Atomic units ($\hbar = e = m_{\rme} = a_0 = 1$) are used throughout.

\section{Calculating spectra}
\label{calculating spectra}

\begin{figure}
 \begin{center}
   \includegraphics[width=\columnwidth]{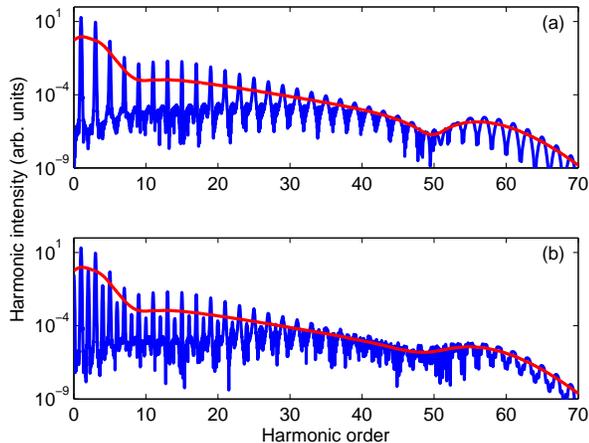}
 \end{center}
 \caption{(Color online) Parallel components of HHG spectra from CO oriented at (a) $\theta = 90^{\circ}$ and (b) $\theta = 86^{\circ}$ to the laser polarization axis. The smooth (red) curves are obtained by averaging the spectra over several harmonic orders. The $800$~nm driving laser pulse has a peak intensity of $4 \times 10^{14}$~W/cm$^2$. Short trajectories are selected using a window function, and the Stark effect is not included.}
 \label{oriented minimum}
\end{figure}

We use CO to illustrate various aspects of HHG from polar molecules. For this proof-of-principle study we restrict ourselves to including the highest occupied molecular orbital (HOMO), ignoring ionization from lower-lying states~\cite{smirnova:063601, smirnova:972, worner:233904, faria:043409} as well as recombination into excited bound states~\cite{han:225601}. The HOMO is a $3\sigma$ orbital, with vertical ionization potential $I_p = 14.014$~eV, a permanent dipole $\mu = 1.1287$~au pointing from C to O, and static polarizabilities $\alpha_{\parallel} = 3.2332$~au and $\alpha_{\perp} = 2.7668$~au parallel and perpendicular to the internuclear axis~\cite{etches:155602}. The nuclei are frozen to the equilibrium distance $\rmR = 1.1283$~\AA. The wave function is obtained using a triple zeta valence basis set in the quantum chemistry software package GAMESS--US~\cite{schmidt:1347}.

The driving pulse is chosen with a linear polarization along the $x$ axis, peak intensity $4 \times 10^{14}$~W/cm$^2$, and a wavelength of 800~nm. The electric field has a $\cos^2$ envelope with 10 optical cycles FWHM. The carrier envelope phase delay has no impact on our results for such long pulses, and is set to zero.

The spectrum $S_{\bfn}(\omega)$ of the harmonic component along a given direction $\bfn$ from a single molecule is given by
\begin{align}
  S_{\bfn}(\theta; \omega) = \left| \mathrm{A}_{\bfn}(\theta; \omega) \right|^2,
\end{align}
where $\mathrm{A}_{\bfn}(\theta; \omega)$ is the projection onto $\bfn$ of the Fourier transform of the time-dependent expectation value of the dipole velocity operator $\braket{\hat{\bfv}_{\mathrm{dip}}(\theta; t)}$~\cite{baggesen}. Due to the cylindrical symmetry of $\sigma$ orbitals, the molecular orientation is defined entirely by the angle $\theta$ between the laser polarization axis and the nuclear displacement $\bfR = \bfR_2 - \bfR_1$. Reference~\cite{etches:155602} explains how to calculate $\braket{\hat{\bfv}_{\mathrm{dip}}(\theta; t)}$ using an extended Lewenstein model that includes the time-dependent Stark shift felt by the molecular orbitals. Short trajectories are selected with a window function, which suppresses trajectories that have excursion times longer than approximately two thirds of an optical cycle.

\begin{figure}
 \begin{center}
   \includegraphics[width=\columnwidth]{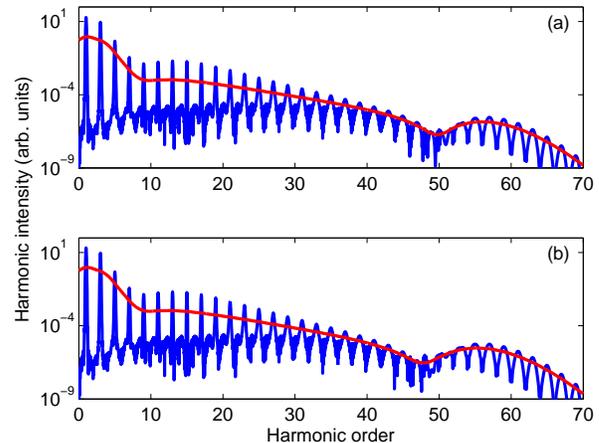}
 \end{center}
 \caption{(Color online) Parallel components of HHG spectra from CO \emph{aligned} at (a) $\theta = 90^{\circ}$ and (b) $\theta = 86^{\circ}$ to the laser polarization axis. All other parameters are the same as those in Fig.~\ref{oriented minimum}.}
 \label{aligned minimum}
\end{figure}

Spectra for CO at two different orientations are plotted in Figure~\ref{oriented minimum}. Figure~\ref{oriented minimum}(a) shows a clear two-center interference minimum from CO oriented perpendicular to the laser polarization axis. However, as soon as the molecule is oriented slightly away from perpendicular in Fig.~\ref{oriented minimum}(b), the minimum becomes less pronounced. Also, Fig.~\ref{oriented minimum}(b) contains even harmonics due to the broken inversion symmetry of the target molecule~\cite{etches:155602, alon:3743, kamta:011404}.

The spectrum from an aligned molecule is obtained by adding the contribution from opposite orientations coherently~\cite{madsen:043419, madsen:035401}
\begin{align}
S_{\bfn, \mathrm{aligned}}(\theta; \omega) = \left| \mathrm{A}_{\bfn}(\theta; \omega) + \mathrm{A}_{\bfn}(\theta + \pi; \omega)\right|^2.
  \label{coherent}
\end{align}
The effect of alignment is seen in Fig.~\ref{aligned minimum}. The spectra of aligned and oriented CO are the same in the perpendicular geometry, but at 86$^{\circ}$ they differ substantially. The first observation is that the spectrum in Fig.~\ref{aligned minimum}(b) contains only odd harmonics, which is due to the inversion symmetry of the aligned target~\cite{etches:155602, alon:3743, kamta:011404}. The second observation is that the interference minimum is much clearer than for oriented CO.

\section{Two-center minimum in aligned molecules}
\label{aligned molecules}

We now present a simple model of the interference minimum in Fig.~\ref{aligned minimum}, and show that it can be thought of either as an interference between recombination at different atoms, or as an interference between opposite orientations.

\begin{figure}
 \begin{center}
   \includegraphics[width=\columnwidth]{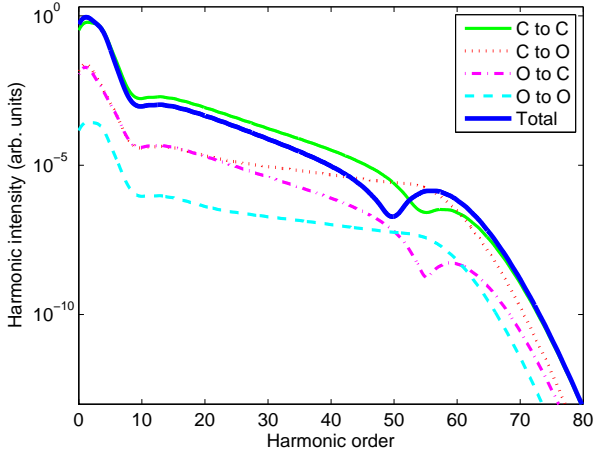}
 \end{center}
 \caption{(Color online) Parallel components of the spectra associated with the four trajectories of CO oriented perpendicular to the laser polarization axis. The molecular spectrum is the coherent sum of all four contributions, and is dominated by trajectories that start with ionization at the carbon atom. Short trajectories are selected, and the Stark effect is not included.}
 \label{trajectories}
\end{figure}

It is a common assumption that two-center interference minima are completely determined by the recombination step into the ground state orbital. This assumption needs to be justified for polar molecules due to the different classes of electron trajectories in the continuum~\cite{chirila:023410, faria:043407, etches:013409}. There are four different trajectories for CO, each one starting at either C or O, and recombining at either C or O. The contribution from each class of trajectories is plotted in Fig.~\ref{trajectories} for CO oriented perpendicular to the laser polarization axis. Trajectories starting at the carbon atom completely dominate the spectrum due to a larger ionization dipole matrix element at carbon. Removing the trajectories starting at oxygen has very little impact on the shape and position of the interference minimum, showing that the interference is not caused by the ionization step. For a detailed discussion of trajectories beginning and ending on the two different atoms, see Refs.~\cite{chirila:023410, faria:043407, etches:013409}.

The trajectory contributions in Fig.~\ref{trajectories} are obtained using the extended stationary-phase approximation~\cite{chirila:023410}, in which the trajectories going from one atom to another acquire an extra component to the momentum. Using the conventional stationary-phase approximation, and thus neglecting this drift momentum, has a negligible effect on the shape and position of the interference minimum. Also, short trajectories were selected using a window function, ruling out interference between long and short trajectories. These observations show that continuum dynamics is not responsible for the interference minimum either.

We now turn to the recombination step, having ruled out interferences in the continuum as well as in the ionization step. First the oriented molecular orbital $\ket{\psi, \theta}$ is expanded on linear combinations of atomic orbitals (LCAO) $\ket{\psi_n, \theta}$
\begin{align}
  \Ket{\psi, \theta} = \sum_n c_n \Ket{\psi_n, \theta}.
  \label{expansion}
\end{align}
The sum is over atomic centers. The recombination dipole velocity matrix element between $\ket{\psi, \theta}$ and a plane wave $\ket{\bfk}$ then takes the form
\begin{align}
    \Braket{\bfk | \hat{\bfv}_{\mathrm{dip}} | \psi, \theta} & = \sum_n \rme^{\rmi \left( \bfk \cdot \bfR_n + \phi_n(\bfk, \theta) \right)} \nonumber \\
    & \phantom{= \sum} \times \left| c_n \Braket{\bfk | \hat{\bfv}_{\mathrm{dip}} | \psi_n, \theta } \right|.
 \label{polar}
\end{align}
The calculation is given in Appendix~\ref{derivations}. The phase $\bfk \cdot \bfR_n$ is due to the atomic position $\bfR_n$, and $\phi_n(\bfk,\theta)$ is an intrinsic recombination phase relating to the LCAO centered on atom $n$. Equation~\eqref{polar} shows that the harmonic emission from different atoms interfere, causing maximal destructive interference at a particular momentum $\bfk$ for a fixed angle $\theta$.

The key to understanding the interference in aligned polar molecules is to realize that the dipole velocity matrix elements of opposite orientations are closely related
\begin{align}
  \Braket{\bfk | \hat{\bfv}_{\mathrm{dip}} | \psi, \theta + \pi} & = \Braket{\bfk | \hat{\bfv}_{\mathrm{dip}} | \psi, \theta}^*.
\label{flipped matrix element}
\end{align}
The calculation is given in Appendix~\ref{derivations}. Together, Eqs.~\eqref{coherent}, \eqref{polar} and \eqref{flipped matrix element} show that the recombination dipole velocity matrix element of an aligned molecule is given by 
\begin{align}
  \Braket{\bfk | \hat{\bfv}_{\mathrm{dip}} | \psi, \theta}_{\mathrm{aligned}} & = \mathrm{Re} \left( \Braket{\bfk | \hat{\bfv}_{\mathrm{dip}} | \psi, \theta} \right) \label{real expectation}\\
  & = \sum_n \cos \left( \bfk \cdot \bfR_n + \phi_n(\bfk, \theta) \right) \nonumber \\
  & \phantom{ = \sum} \; \times \left| c_n \Braket{\bfk | \hat{\bfv}_{\mathrm{dip}} | \psi_n, \theta } \right|.
  \label{aligned expectation}
\end{align}

The position of the two-center minimum is determined by the value of $\bfk$ that minimizes the absolute value of Eq.~\eqref{aligned expectation}. The simplest approach would be to ignore the recombination strengths $\left| c_n \braket{\bfk | \hat{\bfv}_{\mathrm{dip}} | \psi_n, \theta } \right|$, and claim that harmonic emission from one atom cancels out the harmonic emission from another when
\begin{align}
  \rmk \rmR \cos(\theta) + \phi_2(\bfk, \theta) - \phi_1(\bfk, \theta) = (2m + 1) \pi
  \label{crappy prediction}
\end{align}
for any integer $m$. In this approximation, the same result is obtained for oriented molecules using Eq.~\eqref{polar}, and was first presented in Ref.~\cite{zhu:137866}. However, the absolute values of Eqs.~\eqref{polar} and~\eqref{aligned expectation} do not generally have a minimum at the same location, nor can they be predicted entirely from recombination phases. The reason for this is that the recombination strengths in Eqs.~\eqref{polar} and~\eqref{aligned expectation} generally have different $\rmk$ dependences at different atomic centers. The problem disappears in the case of homonuclear molecules, where the matrix norms are identical.

\begin{figure}
 \begin{center}
   \includegraphics[width=\columnwidth]{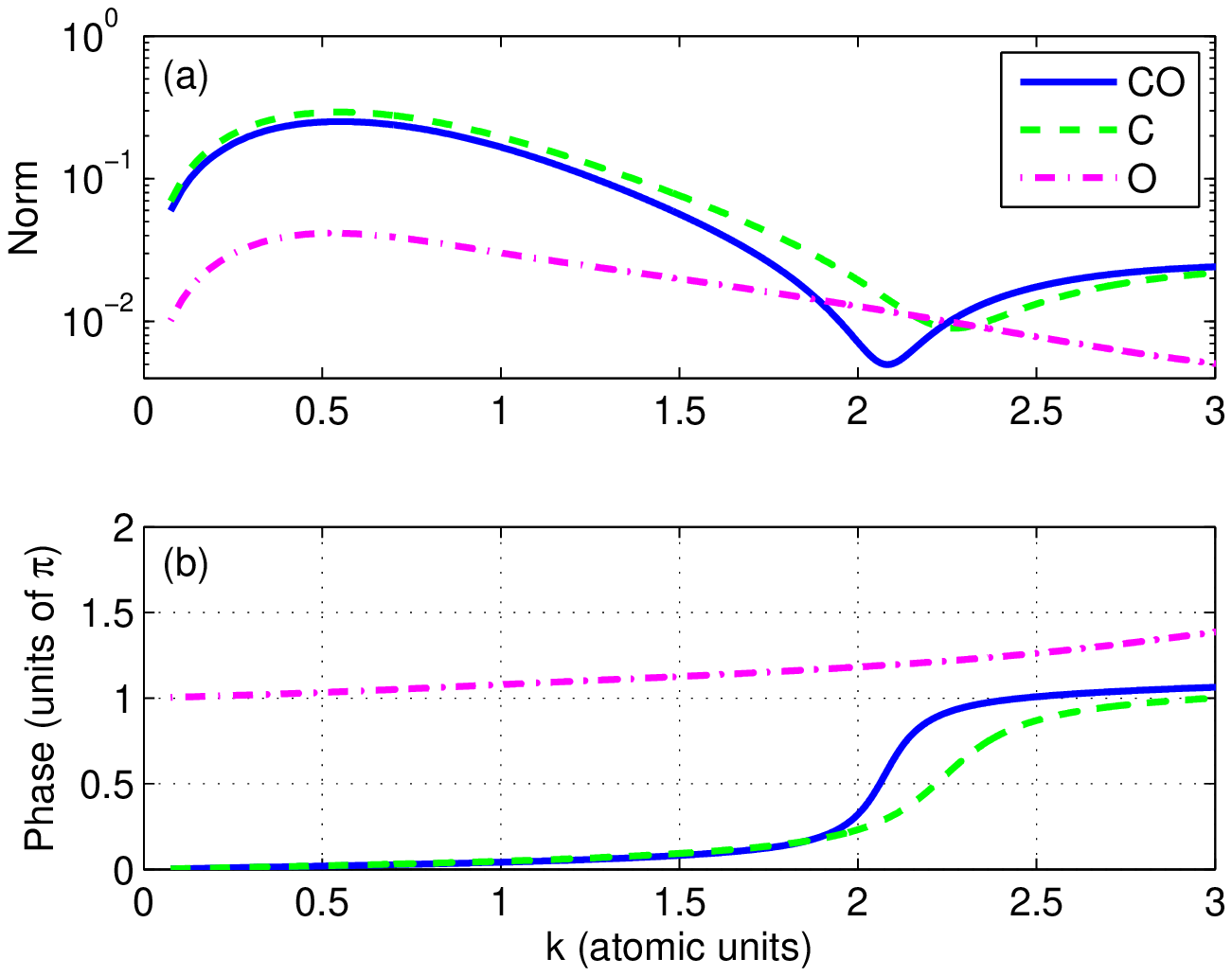}
 \end{center}
 \caption{(Color online) (a) Norms of the dipole velocity matrix elements for CO oriented at $\theta = 86^{\circ}$ to $\bfk$. The full (blue) curve is the matrix element with the entire HOMO $\braket{\bfk | \hat{\bfv}_{\mathrm{dip}} | \psi, 86^{\circ}}$, the dashed (green) curve is the matrix element with the LCAO on carbon $\braket{\bfk | \hat{\bfv}_{\mathrm{dip}} | \psi_1, 86^{\circ}}$, and the dash-dotted (magenta) the matrix element with the LCAO on oxygen $\braket{\bfk | \hat{\bfv}_{\mathrm{dip}} | \psi_2, 86^{\circ}}$. (b) Phases of the same matrix elements. The dashed and dash-dotted curves are given by $\bfk \cdot \bfR_n + \phi_n(\bfk, \theta)$.}
\label{vMatrixNorms}
\end{figure}

The following example illustrates our point. Figure~\ref{vMatrixNorms} shows the relevant recombination dipole velocity matrix elements for CO oriented at $\theta = 86^{\circ}$. The full (blue) curves are the norm and the phase of the total molecular matrix element of CO. The dashed (green) curves are the norm and phase of the matrix element of the LCAO on carbon, and the dash-dotted (magenta) curves on oxygen. The full (blue) curve in Fig.~\ref{vMatrixNorms}(a) has a minimum at $\rmk = 2.08$~au. According to Eq.~\eqref{crappy prediction}, the minimum should be located at the momentum $\rmk$ for which the phase difference between the carbon and oxygen matrix elements in Fig.~\ref{vMatrixNorms}(b) is $\pi$. This gives a predicted minimum at $\rmk = 1.86$~au, which is in poor agreement with the actual position.

The interference minimum in aligned polar molecules can also be thought of as an interference between opposite orientations. Equation~\eqref{real expectation} shows that the interference minimum appears whenever the real part of the recombination matrix element of the total molecular orbital $\braket{\bfk | \hat{\bfv}_{\mathrm{dip}} | \psi, \theta}$ is zero 
\begin{align}
  \mathrm{Re} \left( \Braket{\bfk | \hat{\bfv}_{\mathrm{dip}} | \psi, \theta} \right) = 0.
  \label{good prediction}
\end{align}
According to the solid (blue) curve in Fig.~\ref{vMatrixNorms}(b), this happens at $\rmk = 2.06$~au. Using the relation $\hbar \omega = 0.5 \rmk^2 + I_p$ to convert into photon energy, this corresponds to a minimum at $\omega = 46.3\omega_0$. The actual position of the minimum is $47.8 \omega_0$, as determined by the averaged spectrum in Fig.~\ref{aligned minimum}. For comparison, Eq.~\eqref{crappy prediction} yields a prediction of $37.5\omega_0$. Equation~\eqref{good prediction} is clearly more precise at predicting the minimum position than Eq.~\eqref{crappy prediction}. 

Now we are also able to understand why the interference minimum is stronger for aligned CO in Fig.~\ref{aligned minimum}(b) than for oriented CO in Fig.~\ref{oriented minimum}(b). The strength of the interference minimum in oriented CO is determined by the minimal value of the matrix element norm. As illustrated in Fig.~\ref{vMatrixNorms}, this minimal value is not necessarily zero. For aligned CO only the real part has to be zero, which \emph{is} obtained exactly, hence two-center interference minima are generally expected to be stronger for aligned than for oriented polar molecules.

Equation~\eqref{good prediction} is more precise than Eq.~\eqref{crappy prediction} because no assumptions are made on the recombination strengths at different atomic centers. One might argue that it would be more natural to use Eq.~\eqref{aligned expectation} directly, and read off the minimum in the matrix element norm of the \emph{aligned} molecule rather than focus on the matrix element phase of the \emph{oriented} molecule. However, we show in Sec.~\ref{Stark shift} that the influence of the Stark effect on the minimum position is more easily understood in terms of the recombination phase of the oriented molecule.

\section{Importance of the Stark effect}
\label{Stark shift}

\begin{figure}
 \begin{center}
   \includegraphics[width=\columnwidth]{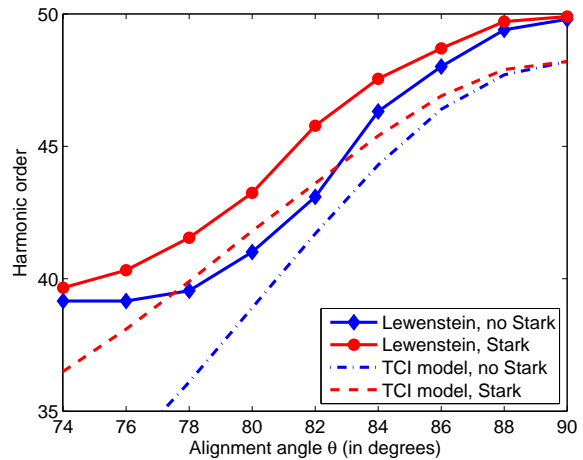}
 \end{center}
 \caption{(Color online) Position of the interference minimum (of the smoothed spectrum) from aligned CO with and without including the Stark effect in the Lewenstein model. Short trajectories have been selected. Also plotted are the predicted minima positions using a two-center interference (TCI) model based on Eq.~\eqref{good prediction}}
 \label{minima positions}
\end{figure}

Equation~\eqref{good prediction} shows that the interference minimum from an aligned polar molecule is expected to depend on the phase of the recombination matrix element of the \emph{oriented} molecule. This begs the question whether phases from other steps of the HHG process influence the final position of the minimum. Figure~\ref{minima positions} shows the position of the interference minimum in aligned CO calculated using an extended Lewenstein model~\cite{etches:155602} with the pulse parameters from Sec.~\ref{calculating spectra}. Including the Stark effect to first order is seen to shift the interference minimum by as much as three harmonic orders. Including the second-order Stark shift has negligible influence on the present spectra.

We propose that the shift of the interference minimum can be understood in terms of the first-order Stark shift that the polar HOMO experiences due to the driving pulse. An electron trajectory starting with ionization at time $t'$, and ending with recombination at time $t$, will accumulate a first-order Stark phase~\cite{etches:155602} given by
\begin{align}
  \Phi_{\mathrm{Stark}} & = \int_{t'}^t  \boldsymbol{\mu} \cdot \bfF(t'') \rmd t'' \\
  & = - \mu \cos(\theta) \left[ \rmA(t) - \rmA(t') \right].
  \label{accumulated Stark}
\end{align}
$\bfA(t)$ is the vector potential of the driving field $\bfF(t) = -\partial_t \bfA(t)$, which is linearly polarized along the $x$ axis. 

\begin{figure}
 \begin{center}
   \includegraphics[width=\columnwidth]{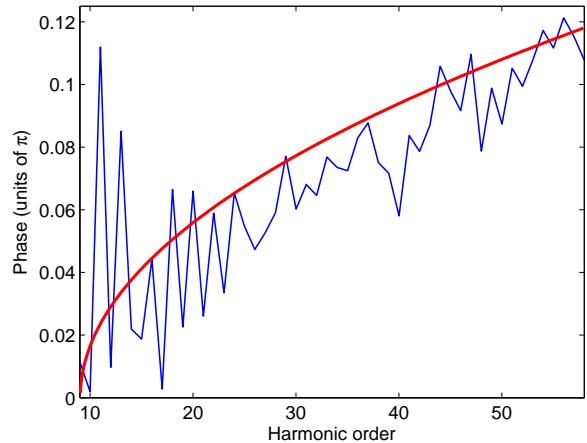}
 \end{center}
 \caption{(Color online) Estimates of the Stark phase accumulated in the continuum for CO oriented at $\theta = 82^{\circ}$, and using the pulse parameters from Sec.~\ref{aligned molecules}. The smooth (red) curve is calculated from Eq.~\eqref{accumulated Stark} based on classical trajectories to map ionization and recombination times to harmonic order. The irregular (blue) curve is the phase difference between harmonics calculated with and without including the Stark effect in the Lewenstein model. To avoid cancellation from neighbouring half-cycles, ionization is artificially restricted to every other half-cycle (see text).}
 \label{Stark phase}
\end{figure}

The accumulated Stark phase in Eq.~\eqref{accumulated Stark} can be estimated using classical trajectories from the three-step model~\cite{krause:3535, corkum:1994}. In short, we consider a single cycle of the electric field, and calculate the return time $t$ and return energy $E_{\mathrm{kin}}$ of a classic electron being liberated at time $t'$ with zero kinetic energy. The accumulated phase is entirely determined by $t$, $t'$ and $\theta$ through Eq.~\eqref{accumulated Stark}, and mapped onto photon energy using $\hbar \omega = E_{\mathrm{kin}} + I_p$. The result is plotted as a smooth (red) curve in Fig.~\ref{Stark phase} for $\theta = 82^{\circ}$. The predicted Stark phase only depends on $\theta$ through an overall factor of $\cos(\theta)$.

An alternative estimate based on the Lewenstein model is obtained by calculating the harmonic spectra from oriented CO with and without including the Stark effect. As neighbouring half-cycles experience opposite signs of the first-order Stark shift, we restrict the ionization step to times $t'$ at which the $x$ component of the electric field $\bfF(t')$ is positive. This singles out continuum trajectories that share the same sign of the Stark shift. Short trajectories are selected using a window function. The difference in harmonic phase between the two calculations is plotted in Fig.~\ref{Stark phase} as an irregular (blue) curve. The agreement with Eq.~\eqref{accumulated Stark} is quite good, especially when keeping in mind the simplicity of the model, and the fact that there are no fitting parameters whatsoever. 

The amount by which the Stark effect shifts the interference minimum can be estimated using Eqs.~\eqref{good prediction} and~\eqref{accumulated Stark}. The interference minimum appears at the momentum $\rmk$ where the real part of the recombination matrix element is zero. For CO oriented at $\theta = 86^{\circ}$ in Fig.~\ref{vMatrixNorms}(b), this happens at $\rmk = 2.06$~au, corresponding to $\omega = 46.3\omega_0$. The accumulated phase from Eq.~\eqref{accumulated Stark} is on the order of $0.05\pi$, which is enough to shift the interference condition to $\rmk = 2.08$~au, corresponding to $\omega = 47.0\omega_0$. When $\theta$ is decreased, the phase shift increases due to the $\cos(\theta)$ scaling in Eq.~\eqref{accumulated Stark}.

\begin{figure}
 \begin{center}
   \includegraphics[width=\columnwidth]{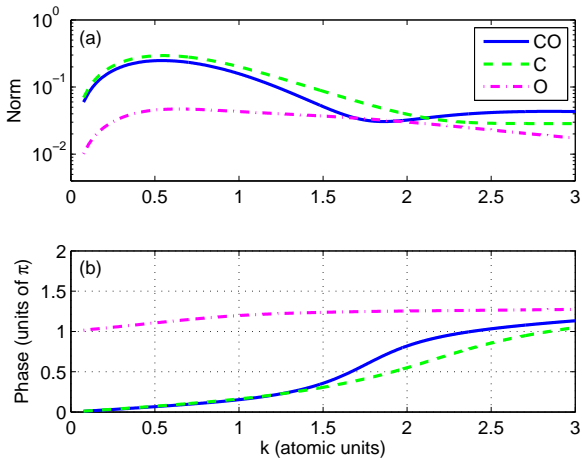}
 \end{center}
 \caption{(Color online) Same as Fig.~\ref{vMatrixNorms}, but with $\theta = 76^{\circ}$.}
\label{vMatrixNorms76}
\end{figure}

Combining Eqs.~\eqref{good prediction} and~\eqref{accumulated Stark} gives the two-center interference predictions plotted in Fig.~\ref{minima positions}. At perpendicular alignment the first-order Stark shift is exactly zero, and the minima coincide. The two-center model is seen to be off by two harmonic orders as compared with the Lewenstein calculations. As the alignment angle is decreased, the two-center interference model adequately describes the shift of the minimum, down to about $\theta = 80^{\circ}$, after which the Lewenstein minimum washes out and finally vanishes. This weakening of the interference minimum at smaller alignment angles is related to a slower phase variation of the total recombination dipole velocity matrix element, illustrated in Fig.~\ref{vMatrixNorms76}(b) as a full (blue) curve for $\theta = 76^{\circ}$. 

\section{Conclusions}
\label{conclusions}

We extend a model of two-center interference to include the superposition of opposite orientations in aligned polar molecules. Our analysis shows that two-center interference minima can be thought of either as an interference between recombination at different atoms or as an interference between opposite orientations. Using the former interpretation, we get the same estimate for the position of the minimum as for oriented molecules~\cite{zhu:137866}. In this approximation the minimum only depends on the phase difference between recombination at different atoms. However, calculations on CO show that the different recombination strengths are also important in determining the position of the minimum.

Interpreting the minimum as an interference between orientations, we predict the minimum position in aligned molecules in terms of the phase of the recombination dipole velocity matrix element of the \emph{oriented} molecule. This model successfully predicts the minimum position in spectra calculated for aligned CO to within two harmonic orders for alignment angles ranging between 80 and 90 degrees. At smaller alignment angles the observed minimum vanishes.

Inclusion of the Stark effect shifts the observed interference minimum \emph{even though aligned molecules do not posses total permanent dipoles}. We explain this by the fact that the interference minimum is determined by the recombination phase of the \emph{oriented} molecule, which does have a permanent dipole. This recombination phase is modified by the first-order Stark phase that the electron picks up between ionization and recombination. Our model successfully reproduces the shift of the interference minimum due to the Stark effect.

\appendix

\section{}
\label{derivations}

Here we derive Eqs.~\eqref{polar} and~\eqref{flipped matrix element}. The ground state of the oriented molecule is denoted $\ket{\psi, \theta}$, and the corresponding wave function in the laboratory frame $\psi(\theta; \bfr)$. Using the multi-center expansion in Eq.~\eqref{expansion} the total recombination dipole velocity matrix element is
\begin{align}
    \Braket{\bfk | \hat{\bfv}_{\mathrm{dip}} | \psi, \theta} & = \int \psi_{\bfk}^*(\bfr) \hat{\bfv}_{\mathrm{dip}} \psi(\theta; \bfr) \rmd \bfr \\ 
  & = \int \psi_{\bfk}^*(\bfr) \hat{\bfv}_{\mathrm{dip}} \sum_n c_n \psi_n(\theta; \bfr - \bfR_n) \rmd \bfr \\ 
  & = \sum_n c_n \int \rme^{\rmi \bfk \cdot \bfr} \hat{\bfv}_{\mathrm{dip}} \psi_n(\theta; \bfr - \bfR_n) \rmd \bfr \\
  & = \sum_n c_n \rme^{\rmi \bfk \cdot \bfR_n} \int \rme^{\rmi \bfk \cdot \bfr} \hat{\bfv}_{\mathrm{dip}} \psi_n(\theta; \bfr) \rmd \bfr \\
  & = \sum_n c_n \rme^{\rmi \bfk \cdot \bfR_n} \Braket{\bfk | \hat{\bfv}_{\mathrm{dip}} | \psi_n, \theta} \\
  & = \sum_n \rme^{\rmi \left( \bfk \cdot \bfR_n + \phi_n(\bfk, \theta) \right)} \nonumber \\
    & \phantom{= \sum} \times \left| c_n \Braket{\bfk | \hat{\bfv}_{\mathrm{dip}} | \psi_n, \theta } \right|.
\end{align}
In the last step the intrinsic recombination phase $\phi_n(\bfk, \theta)$ of the matrix element $c_n \Braket{\bfk | \hat{\bfv}_{\mathrm{dip}} | \psi_n, \theta }$ is written out explicitly.

Equation~\eqref{flipped matrix element} follows from the observation that the recombination dipole velocity matrix element is given by
\begin{align}
  \Braket{\bfk | \hat{\bfv}_{\mathrm{dip}} | \psi, \theta} & = \bfk \nabla_{\bfk} \tilde{\psi}(\theta; \bfk),
\end{align}
where $\tilde{\psi}(\theta; \bfk)$ is the Fourier transform of the ground state wave function. Flipping the orientation of the molecule is the same as adding $\pi$ to $\theta$. We now fix the coordinate system such that the laser polarization is along the $x$ axis and the molecular axis confined to the $xy$ plane. All molecular orbitals can then be chosen to be real, and either even or odd in $z$. Orbitals that are odd in $z$ do not contribute to the total HHG spectrum~\cite{madsen:043419}, which leaves orbitals for which adding $\pi$ to $\theta$ is the same as replacing the wave function with $\psi(\theta; -x, -y, z) = \psi(\theta; -x, -y, -z)$.

The recombination dipole velocity matrix element of a flipped molecule is then
\begin{align}
  \Braket{\bfk | \hat{\bfv}_{\mathrm{dip}} | \psi, \theta + \pi} & = \bfk \nabla_{\bfk} \tilde{\psi}(\theta + \pi; \bfk) \\
  & = \bfk \nabla_{\bfk} \tilde{\psi}(\theta; -\bfk) \\
  & = \bfk \nabla_{\bfk} \tilde{\psi}^*(\theta; \bfk) \\
  & = \Braket{\bfk | \hat{\bfv}_{\mathrm{dip}} | \psi, \theta}^*.
\end{align}
In the above we use the behaviour of the Fourier transform under coordinate inversion as well as a property of Fourier transforms of real functions.

\section*{Acknowledgements}

This work was supported by the Danish National Research Council (Grant No.~10-85430), and the National Science Foundation under Grant No.~PHY-1019071.

\bibliography{bibfile}

\begin{thebibliography}{30}
\expandafter\ifx\csname natexlab\endcsname\relax\def\natexlab#1{#1}\fi
\expandafter\ifx\csname bibnamefont\endcsname\relax
  \def\bibnamefont#1{#1}\fi
\expandafter\ifx\csname bibfnamefont\endcsname\relax
  \def\bibfnamefont#1{#1}\fi
\expandafter\ifx\csname citenamefont\endcsname\relax
  \def\citenamefont#1{#1}\fi
\expandafter\ifx\csname url\endcsname\relax
  \def\url#1{\texttt{#1}}\fi
\expandafter\ifx\csname urlprefix\endcsname\relax\def\urlprefix{URL }\fi
\providecommand{\bibinfo}[2]{#2}
\providecommand{\eprint}[2][]{\url{#2}}

\bibitem[{\citenamefont{Lein et~al.}(2002{\natexlab{a}})\citenamefont{Lein,
  Hay, Velotta, Marangos, and Knight}}]{lein:023805}
\bibinfo{author}{\bibfnamefont{M.}~\bibnamefont{Lein}},
  \bibinfo{author}{\bibfnamefont{N.}~\bibnamefont{Hay}},
  \bibinfo{author}{\bibfnamefont{R.}~\bibnamefont{Velotta}},
  \bibinfo{author}{\bibfnamefont{J.~P.} \bibnamefont{Marangos}},
  \bibnamefont{and} \bibinfo{author}{\bibfnamefont{P.~L.}
  \bibnamefont{Knight}}, \bibinfo{journal}{Phys. Rev. A}
  \textbf{\bibinfo{volume}{66}}, \bibinfo{pages}{023805}
  (\bibinfo{year}{2002}{\natexlab{a}}).

\bibitem[{\citenamefont{Lein et~al.}(2002{\natexlab{b}})\citenamefont{Lein,
  Hay, Velotta, Marangos, and Knight}}]{lein:183903}
\bibinfo{author}{\bibfnamefont{M.}~\bibnamefont{Lein}},
  \bibinfo{author}{\bibfnamefont{N.}~\bibnamefont{Hay}},
  \bibinfo{author}{\bibfnamefont{R.}~\bibnamefont{Velotta}},
  \bibinfo{author}{\bibfnamefont{J.~P.} \bibnamefont{Marangos}},
  \bibnamefont{and} \bibinfo{author}{\bibfnamefont{P.~L.}
  \bibnamefont{Knight}}, \bibinfo{journal}{Phys. Rev. Lett.}
  \textbf{\bibinfo{volume}{88}}, \bibinfo{pages}{183903}
  (\bibinfo{year}{2002}{\natexlab{b}}).

\bibitem[{\citenamefont{Krause et~al.}(1992)\citenamefont{Krause, Schafer, and
  Kulander}}]{krause:3535}
\bibinfo{author}{\bibfnamefont{J.~L.} \bibnamefont{Krause}},
  \bibinfo{author}{\bibfnamefont{K.~J.} \bibnamefont{Schafer}},
  \bibnamefont{and} \bibinfo{author}{\bibfnamefont{K.~C.}
  \bibnamefont{Kulander}}, \bibinfo{journal}{Phys. Rev. Lett.}
  \textbf{\bibinfo{volume}{68}}, \bibinfo{pages}{3535} (\bibinfo{year}{1992}).

\bibitem[{\citenamefont{Corkum}(1993)}]{corkum:1994}
\bibinfo{author}{\bibfnamefont{P.~B.} \bibnamefont{Corkum}},
  \bibinfo{journal}{Phys. Rev. Lett.} \textbf{\bibinfo{volume}{71}},
  \bibinfo{pages}{1994} (\bibinfo{year}{1993}).

\bibitem[{\citenamefont{Lewenstein et~al.}(1994)\citenamefont{Lewenstein,
  Balcou, Ivanov, L\char39{}Huillier, and Corkum}}]{lewenstein:2117}
\bibinfo{author}{\bibfnamefont{M.}~\bibnamefont{Lewenstein}},
  \bibinfo{author}{\bibfnamefont{P.}~\bibnamefont{Balcou}},
  \bibinfo{author}{\bibfnamefont{M.~Y.} \bibnamefont{Ivanov}},
  \bibinfo{author}{\bibfnamefont{A.}~\bibnamefont{L\char39{}Huillier}},
  \bibnamefont{and} \bibinfo{author}{\bibfnamefont{P.~B.}
  \bibnamefont{Corkum}}, \bibinfo{journal}{Phys. Rev. A}
  \textbf{\bibinfo{volume}{49}}, \bibinfo{pages}{2117} (\bibinfo{year}{1994}).

\bibitem[{\citenamefont{Zhu et~al.}(2011)\citenamefont{Zhu, Zhang, Hong, Lan,
  and Lu}}]{zhu:137866}
\bibinfo{author}{\bibfnamefont{X.}~\bibnamefont{Zhu}},
  \bibinfo{author}{\bibfnamefont{Q.}~\bibnamefont{Zhang}},
  \bibinfo{author}{\bibfnamefont{W.}~\bibnamefont{Hong}},
  \bibinfo{author}{\bibfnamefont{P.}~\bibnamefont{Lan}}, \bibnamefont{and}
  \bibinfo{author}{\bibfnamefont{P.}~\bibnamefont{Lu}},
  \bibinfo{journal}{Optics Express} \textbf{\bibinfo{volume}{19}},
  \bibinfo{pages}{137866} (\bibinfo{year}{2011}).

\bibitem[{\citenamefont{Stapelfeldt and Seideman}(2003)}]{stapelfeldt:543}
\bibinfo{author}{\bibfnamefont{H.}~\bibnamefont{Stapelfeldt}} \bibnamefont{and}
  \bibinfo{author}{\bibfnamefont{T.}~\bibnamefont{Seideman}},
  \bibinfo{journal}{Rev. Mod. Phys.} \textbf{\bibinfo{volume}{75}},
  \bibinfo{pages}{543} (\bibinfo{year}{2003}).

\bibitem[{\citenamefont{Od\v{z}ak and Milo\v{s}evi\'{c}}(2009)}]{odzak:023414}
\bibinfo{author}{\bibfnamefont{S.}~\bibnamefont{Od\v{z}ak}} \bibnamefont{and}
  \bibinfo{author}{\bibfnamefont{D.~B.} \bibnamefont{Milo\v{s}evi\'{c}}},
  \bibinfo{journal}{Phys. Rev. A} \textbf{\bibinfo{volume}{79}},
  \bibinfo{pages}{023414} (\bibinfo{year}{2009}).

\bibitem[{\citenamefont{Augstein and
  de~Morisson~Faria}(2011)}]{augstein:055601}
\bibinfo{author}{\bibfnamefont{B.~B.} \bibnamefont{Augstein}} \bibnamefont{and}
  \bibinfo{author}{\bibfnamefont{C.~F.} \bibnamefont{de~Morisson~Faria}},
  \bibinfo{journal}{J. Phys. B} \textbf{\bibinfo{volume}{44}},
  \bibinfo{pages}{055601} (\bibinfo{year}{2011}).

\bibitem[{\citenamefont{Etches and Madsen}(2010)}]{etches:155602}
\bibinfo{author}{\bibfnamefont{A.}~\bibnamefont{Etches}} \bibnamefont{and}
  \bibinfo{author}{\bibfnamefont{L.~B.} \bibnamefont{Madsen}},
  \bibinfo{journal}{J. Phys. B.} \textbf{\bibinfo{volume}{43}},
  \bibinfo{pages}{155602} (\bibinfo{year}{2010}).

\bibitem[{\citenamefont{Holmegaard et~al.}(2010)\citenamefont{Holmegaard,
  Hansen, Kalhoj, Kragh, Stapelfeldt, Filsinger, Kupper, Meijer, Dimitrovski,
  Abu-samha et~al.}}]{holmegaard:428}
\bibinfo{author}{\bibfnamefont{L.}~\bibnamefont{Holmegaard}},
  \bibinfo{author}{\bibfnamefont{J.~L.} \bibnamefont{Hansen}},
  \bibinfo{author}{\bibfnamefont{L.}~\bibnamefont{Kalhoj}},
  \bibinfo{author}{\bibfnamefont{S.~L.} \bibnamefont{Kragh}},
  \bibinfo{author}{\bibfnamefont{H.}~\bibnamefont{Stapelfeldt}},
  \bibinfo{author}{\bibfnamefont{F.}~\bibnamefont{Filsinger}},
  \bibinfo{author}{\bibfnamefont{J.}~\bibnamefont{Kupper}},
  \bibinfo{author}{\bibfnamefont{G.}~\bibnamefont{Meijer}},
  \bibinfo{author}{\bibfnamefont{D.}~\bibnamefont{Dimitrovski}},
  \bibinfo{author}{\bibfnamefont{M.}~\bibnamefont{Abu-samha}},
  \bibnamefont{et~al.}, \bibinfo{journal}{Nature Physics}
  \textbf{\bibinfo{volume}{6}}, \bibinfo{pages}{428} (\bibinfo{year}{2010}).

\bibitem[{\citenamefont{Dimitrovski et~al.}(2011)\citenamefont{Dimitrovski,
  Abu-samha, Madsen, Filsinger, Meijer, K\"upper, Holmegaard, Kalh\o{}j,
  Nielsen, and Stapelfeldt}}]{dimitrovski:023405}
\bibinfo{author}{\bibfnamefont{D.}~\bibnamefont{Dimitrovski}},
  \bibinfo{author}{\bibfnamefont{M.}~\bibnamefont{Abu-samha}},
  \bibinfo{author}{\bibfnamefont{L.~B.} \bibnamefont{Madsen}},
  \bibinfo{author}{\bibfnamefont{F.}~\bibnamefont{Filsinger}},
  \bibinfo{author}{\bibfnamefont{G.}~\bibnamefont{Meijer}},
  \bibinfo{author}{\bibfnamefont{J.}~\bibnamefont{K\"upper}},
  \bibinfo{author}{\bibfnamefont{L.}~\bibnamefont{Holmegaard}},
  \bibinfo{author}{\bibfnamefont{L.}~\bibnamefont{Kalh\o{}j}},
  \bibinfo{author}{\bibfnamefont{J.~H.} \bibnamefont{Nielsen}},
  \bibnamefont{and}
  \bibinfo{author}{\bibfnamefont{H.}~\bibnamefont{Stapelfeldt}},
  \bibinfo{journal}{Phys. Rev. A} \textbf{\bibinfo{volume}{83}},
  \bibinfo{pages}{023405} (\bibinfo{year}{2011}).

\bibitem[{\citenamefont{Hansen et~al.}(2011)\citenamefont{Hansen, Holmegaard,
  Kalh\o{}j, Kragh, Stapelfeldt, Filsinger, Meijer, K\"upper, Dimitrovski,
  Abu-samha et~al.}}]{hansen:023406}
\bibinfo{author}{\bibfnamefont{J.~L.} \bibnamefont{Hansen}},
  \bibinfo{author}{\bibfnamefont{L.}~\bibnamefont{Holmegaard}},
  \bibinfo{author}{\bibfnamefont{L.}~\bibnamefont{Kalh\o{}j}},
  \bibinfo{author}{\bibfnamefont{S.~L.} \bibnamefont{Kragh}},
  \bibinfo{author}{\bibfnamefont{H.}~\bibnamefont{Stapelfeldt}},
  \bibinfo{author}{\bibfnamefont{F.}~\bibnamefont{Filsinger}},
  \bibinfo{author}{\bibfnamefont{G.}~\bibnamefont{Meijer}},
  \bibinfo{author}{\bibfnamefont{J.}~\bibnamefont{K\"upper}},
  \bibinfo{author}{\bibfnamefont{D.}~\bibnamefont{Dimitrovski}},
  \bibinfo{author}{\bibfnamefont{M.}~\bibnamefont{Abu-samha}},
  \bibnamefont{et~al.}, \bibinfo{journal}{Phys. Rev. A}
  \textbf{\bibinfo{volume}{83}}, \bibinfo{pages}{023406}
  (\bibinfo{year}{2011}).

\bibitem[{\citenamefont{Lein}(2011)}]{lein:ModOpt}
\bibinfo{author}{\bibfnamefont{M.}~\bibnamefont{Lein}}, \bibinfo{journal}{J.
  Mod. Opt.} \textbf{\bibinfo{volume}{in print}} (\bibinfo{year}{2011}).

\bibitem[{\citenamefont{Dimitrovski et~al.}(2010)\citenamefont{Dimitrovski,
  Martiny, and Madsen}}]{dimitrovski:053404}
\bibinfo{author}{\bibfnamefont{D.}~\bibnamefont{Dimitrovski}},
  \bibinfo{author}{\bibfnamefont{C.~P.~J.} \bibnamefont{Martiny}},
  \bibnamefont{and} \bibinfo{author}{\bibfnamefont{L.~B.}
  \bibnamefont{Madsen}}, \bibinfo{journal}{Phys. Rev. A}
  \textbf{\bibinfo{volume}{82}}, \bibinfo{pages}{053404}
  (\bibinfo{year}{2010}).

\bibitem[{\citenamefont{Abu-samha and Madsen}(2010)}]{abu-samha:043413}
\bibinfo{author}{\bibfnamefont{M.}~\bibnamefont{Abu-samha}} \bibnamefont{and}
  \bibinfo{author}{\bibfnamefont{L.~B.} \bibnamefont{Madsen}},
  \bibinfo{journal}{Phys. Rev. A} \textbf{\bibinfo{volume}{82}},
  \bibinfo{pages}{043413} (\bibinfo{year}{2010}).

\bibitem[{\citenamefont{Smirnova
  et~al.}(2009{\natexlab{a}})\citenamefont{Smirnova, Patchkovskii, Mairesse,
  Dudovich, Villeneuve, Corkum, and Ivanov}}]{smirnova:063601}
\bibinfo{author}{\bibfnamefont{O.}~\bibnamefont{Smirnova}},
  \bibinfo{author}{\bibfnamefont{S.}~\bibnamefont{Patchkovskii}},
  \bibinfo{author}{\bibfnamefont{Y.}~\bibnamefont{Mairesse}},
  \bibinfo{author}{\bibfnamefont{N.}~\bibnamefont{Dudovich}},
  \bibinfo{author}{\bibfnamefont{D.}~\bibnamefont{Villeneuve}},
  \bibinfo{author}{\bibfnamefont{P.}~\bibnamefont{Corkum}}, \bibnamefont{and}
  \bibinfo{author}{\bibfnamefont{M.~Y.} \bibnamefont{Ivanov}},
  \bibinfo{journal}{Phys. Rev. Lett.} \textbf{\bibinfo{volume}{102}},
  \bibinfo{pages}{063601} (\bibinfo{year}{2009}{\natexlab{a}}).

\bibitem[{\citenamefont{Smirnova
  et~al.}(2009{\natexlab{b}})\citenamefont{Smirnova, Mairesse, Patchkovskii,
  Dudovich, Villeneuve, Corkum, and Ivanov}}]{smirnova:972}
\bibinfo{author}{\bibfnamefont{O.}~\bibnamefont{Smirnova}},
  \bibinfo{author}{\bibfnamefont{Y.}~\bibnamefont{Mairesse}},
  \bibinfo{author}{\bibfnamefont{S.}~\bibnamefont{Patchkovskii}},
  \bibinfo{author}{\bibfnamefont{N.}~\bibnamefont{Dudovich}},
  \bibinfo{author}{\bibfnamefont{D.}~\bibnamefont{Villeneuve}},
  \bibinfo{author}{\bibfnamefont{P.}~\bibnamefont{Corkum}}, \bibnamefont{and}
  \bibinfo{author}{\bibfnamefont{M.}~\bibnamefont{Ivanov}},
  \bibinfo{journal}{Nature} \textbf{\bibinfo{volume}{460}},
  \bibinfo{pages}{972} (\bibinfo{year}{2009}{\natexlab{b}}).

\bibitem[{\citenamefont{W\"orner et~al.}(2010)\citenamefont{W\"orner, Bertrand,
  Hockett, Corkum, and Villeneuve}}]{worner:233904}
\bibinfo{author}{\bibfnamefont{H.~J.} \bibnamefont{W\"orner}},
  \bibinfo{author}{\bibfnamefont{J.~B.} \bibnamefont{Bertrand}},
  \bibinfo{author}{\bibfnamefont{P.}~\bibnamefont{Hockett}},
  \bibinfo{author}{\bibfnamefont{P.~B.} \bibnamefont{Corkum}},
  \bibnamefont{and} \bibinfo{author}{\bibfnamefont{D.~M.}
  \bibnamefont{Villeneuve}}, \bibinfo{journal}{Phys. Rev. Lett.}
  \textbf{\bibinfo{volume}{104}}, \bibinfo{pages}{233904}
  (\bibinfo{year}{2010}).

\bibitem[{\citenamefont{Figueira~de Morisson~Faria and
  Augstein}(2010)}]{faria:043409}
\bibinfo{author}{\bibfnamefont{C.}~\bibnamefont{Figueira~de Morisson~Faria}}
  \bibnamefont{and} \bibinfo{author}{\bibfnamefont{B.~B.}
  \bibnamefont{Augstein}}, \bibinfo{journal}{Phys. Rev. A}
  \textbf{\bibinfo{volume}{81}}, \bibinfo{pages}{043409}
  (\bibinfo{year}{2010}).

\bibitem[{\citenamefont{Han and Madsen}(2010)}]{han:225601}
\bibinfo{author}{\bibfnamefont{Y.-C.} \bibnamefont{Han}} \bibnamefont{and}
  \bibinfo{author}{\bibfnamefont{L.~B.} \bibnamefont{Madsen}},
  \bibinfo{journal}{J. Phys. B} \textbf{\bibinfo{volume}{43}},
  \bibinfo{pages}{225601} (\bibinfo{year}{2010}).

\bibitem[{\citenamefont{Schmidt et~al.}(1993)\citenamefont{Schmidt, Baldridge,
  Boatz, Elbert, Gordon, Jensen, Koseki, Matsunaga, Nguyen, Su
  et~al.}}]{schmidt:1347}
\bibinfo{author}{\bibfnamefont{M.}~\bibnamefont{Schmidt}},
  \bibinfo{author}{\bibfnamefont{K.}~\bibnamefont{Baldridge}},
  \bibinfo{author}{\bibfnamefont{J.}~\bibnamefont{Boatz}},
  \bibinfo{author}{\bibfnamefont{S.}~\bibnamefont{Elbert}},
  \bibinfo{author}{\bibfnamefont{M.}~\bibnamefont{Gordon}},
  \bibinfo{author}{\bibfnamefont{J.}~\bibnamefont{Jensen}},
  \bibinfo{author}{\bibfnamefont{S.}~\bibnamefont{Koseki}},
  \bibinfo{author}{\bibfnamefont{N.}~\bibnamefont{Matsunaga}},
  \bibinfo{author}{\bibfnamefont{K.}~\bibnamefont{Nguyen}},
  \bibinfo{author}{\bibfnamefont{S.}~\bibnamefont{Su}}, \bibnamefont{et~al.},
  \bibinfo{journal}{J. Comput. Chem.} \textbf{\bibinfo{volume}{14}},
  \bibinfo{pages}{1347} (\bibinfo{year}{1993}).

\bibitem[{\citenamefont{Baggesen and Madsen}(2011)}]{baggesen}
\bibinfo{author}{\bibfnamefont{J.~C.} \bibnamefont{Baggesen}} \bibnamefont{and}
  \bibinfo{author}{\bibfnamefont{L.~B.} \bibnamefont{Madsen}},
  \bibinfo{journal}{J. Phys. B} \textbf{\bibinfo{volume}{44}},
  \bibinfo{pages}{115601} (\bibinfo{year}{2011}).

\bibitem[{\citenamefont{Alon et~al.}(1998)\citenamefont{Alon, Averbukh, and
  Moiseyev}}]{alon:3743}
\bibinfo{author}{\bibfnamefont{O.~E.} \bibnamefont{Alon}},
  \bibinfo{author}{\bibfnamefont{V.}~\bibnamefont{Averbukh}}, \bibnamefont{and}
  \bibinfo{author}{\bibfnamefont{N.}~\bibnamefont{Moiseyev}},
  \bibinfo{journal}{Phys. Rev. Lett.} \textbf{\bibinfo{volume}{80}},
  \bibinfo{pages}{3743} (\bibinfo{year}{1998}).

\bibitem[{\citenamefont{Kamta and Bandrauk}(2004)}]{kamta:011404}
\bibinfo{author}{\bibfnamefont{G.~L.} \bibnamefont{Kamta}} \bibnamefont{and}
  \bibinfo{author}{\bibfnamefont{A.~D.} \bibnamefont{Bandrauk}},
  \bibinfo{journal}{Phys. Rev. A} \textbf{\bibinfo{volume}{70}},
  \bibinfo{pages}{011404} (\bibinfo{year}{2004}).

\bibitem[{\citenamefont{Madsen and Madsen}(2007)}]{madsen:043419}
\bibinfo{author}{\bibfnamefont{C.~B.} \bibnamefont{Madsen}} \bibnamefont{and}
  \bibinfo{author}{\bibfnamefont{L.~B.} \bibnamefont{Madsen}},
  \bibinfo{journal}{Phys. Rev. A} \textbf{\bibinfo{volume}{76}},
  \bibinfo{pages}{043419} (\bibinfo{year}{2007}).

\bibitem[{\citenamefont{Madsen et~al.}(2007)\citenamefont{Madsen, Mouritzen,
  Kjeldsen, and Madsen}}]{madsen:035401}
\bibinfo{author}{\bibfnamefont{C.~B.} \bibnamefont{Madsen}},
  \bibinfo{author}{\bibfnamefont{A.~S.} \bibnamefont{Mouritzen}},
  \bibinfo{author}{\bibfnamefont{T.~K.} \bibnamefont{Kjeldsen}},
  \bibnamefont{and} \bibinfo{author}{\bibfnamefont{L.~B.}
  \bibnamefont{Madsen}}, \bibinfo{journal}{Phys. Rev. A}
  \textbf{\bibinfo{volume}{76}}, \bibinfo{pages}{035401}
  (\bibinfo{year}{2007}).

\bibitem[{\citenamefont{Chiril\u{a} and Lein}(2006)}]{chirila:023410}
\bibinfo{author}{\bibfnamefont{C.~C.} \bibnamefont{Chiril\u{a}}}
  \bibnamefont{and} \bibinfo{author}{\bibfnamefont{M.}~\bibnamefont{Lein}},
  \bibinfo{journal}{Phys. Rev. A} \textbf{\bibinfo{volume}{73}},
  \bibinfo{pages}{023410} (\bibinfo{year}{2006}).

\bibitem[{\citenamefont{de~Morisson~Faria}(2007)}]{faria:043407}
\bibinfo{author}{\bibfnamefont{C.~F.} \bibnamefont{de~Morisson~Faria}},
  \bibinfo{journal}{Phys. Rev. A} \textbf{\bibinfo{volume}{76}},
  \bibinfo{pages}{043407} (\bibinfo{year}{2007}).

\bibitem[{\citenamefont{Etches et~al.}(2010)\citenamefont{Etches, Madsen, and
  Madsen}}]{etches:013409}
\bibinfo{author}{\bibfnamefont{A.}~\bibnamefont{Etches}},
  \bibinfo{author}{\bibfnamefont{C.~B.} \bibnamefont{Madsen}},
  \bibnamefont{and} \bibinfo{author}{\bibfnamefont{L.~B.}
  \bibnamefont{Madsen}}, \bibinfo{journal}{Phys. Rev. A}
  \textbf{\bibinfo{volume}{81}}, \bibinfo{pages}{013409}
  (\bibinfo{year}{2010}).

\end{thebibliography}

\end{document}